\documentclass{article}
\usepackage{authblk}
\usepackage{spconf,amsmath,graphicx}

\usepackage{enumitem}
\setlist{nosep, leftmargin=14pt}

\usepackage{mwe} 
\usepackage[T1]{fontenc}
\usepackage{graphicx}
\usepackage{amsmath,amscd}
\usepackage[section]{placeins}
\usepackage{amssymb,array}
\usepackage{amsfonts,latexsym}
\usepackage{graphicx,subfig,wrapfig}
\usepackage{times}
\usepackage{psfrag,epsfig}
\usepackage{verbatim}
\usepackage{tabularx}
\usepackage[pagebackref=true,breaklinks=true,letterpaper=true,colorlinks,bookmarks=false]{hyperref}
\usepackage{cite}
\usepackage{algorithm}
\usepackage{multirow}
\usepackage{caption}
\usepackage{algorithmic}
\usepackage[amsmath,thmmarks]{ntheorem}
\usepackage{listings}
\usepackage{color}
\usepackage{booktabs}

\title{Cas-DiffCom: Cascaded diffusion model for infant longitudinal super-resolution 3D medical image completion}
\name{\parbox{\linewidth}{\centering Lianghu Guo$^{1,\dagger}$\thanks{$\dagger$ indicates co-first authors and * indicates corresponding authors. This work is partially supported by the STI 2030-Major Projects (No. 2022ZD0209000), National Natural Science Foundation of China (No. 62203355), Shanghai Pilot Program for Basic Research - Chinese Academy of Science, Shanghai Branch (No. JCYJ-SHFY-2022-014), Shenzhen Science and Technology Program (No. KCXFZ20211020163408012), and Shanghai Pujiang Program (No. 21PJ1421400).} \quad Tianli Tao$^{1,\dagger}$ \quad Xinyi Cai$^{1,\dagger}$ \quad Zihao Zhu$^{1,\dagger}$ \quad Jiawei Huang$^{1}$\\ \quad Lixuan Zhu$^{1}$ \quad Zhuoyang Gu$^{1}$ \quad Haifeng Tang$^{1}$ \quad Rui Zhou$^{1}$ \quad Siyan Han$^{1}$\\ Yan Liang$^{1}$ \quad Qing Yang$^{1}$ \quad Dinggang Shen$^{1,2,3}$ \quad Han Zhang$^{1,3,*}$}}

\address{$^{1}$ School of Biomedical Engineering, ShanghaiTech University, Shanghai, China \\
    $^{2}$ Shanghai United Imaging Intelligence Co., Ltd, Shanghai, China \\
    $^{3}$ Shanghai Clinical Research and Trial Center, Shanghai, 201210, China}
\begin{document}
%
\maketitle
\begin{abstract}
Early infancy is a rapid and dynamic neurodevelopmental period for behavior and neurocognition. Longitudinal magnetic resonance imaging (MRI) is an effective tool to investigate such a crucial stage by capturing the developmental trajectories of the brain structures. However, longitudinal MRI acquisition always meets a serious data-missing problem due to participant dropout and failed scans, making longitudinal infant brain atlas construction and developmental trajectory delineation quite challenging. Thanks to the development of an AI-based generative model, neuroimage completion has become a powerful technique to retain as much available data as possible. However, current image completion methods usually suffer from inconsistency within each individual subject in the time dimension, compromising the overall quality. To solve this problem, our paper proposed a two-stage cascaded diffusion model, Cas-DiffCom, for dense and longitudinal 3D infant brain MRI completion and super-resolution. We applied our proposed method to the Baby Connectome Project (BCP) dataset. The experiment results validate that Cas-DiffCom achieves both individual consistency and high fidelity in longitudinal infant brain image completion. We further applied the generated infant brain images to two downstream tasks, brain tissue segmentation and developmental trajectory delineation, to declare its task-oriented potential in the neuroscience field. 
\end{abstract}
\begin{keywords}
Medical imaging completion, Diffusion model, Super-resolution, Infant development, MRI
\end{keywords}
\section{Introduction}
\label{sec:intro}

During the initial two years after birth, the human brain undergoes a swift development in its structure and morphology\cite{gilmore2018brainDevelopment}. Longitudinal studies have unique advantages in accounting for infant brain anatomical development and disease progression variations. Densely collecting brain images to construct a longitudinal medical image dataset has drawn extensive attention due to its profound applications in the neuroscience field, such as developmental brain atlas construction, growth charting, brain age and developmental outcome prediction.

However, research on early brain development faces challenges in both image acquisition and analysis due to the low scanning success rate for infants\cite{gilmore2018brainDevelopment}. Moreover, factors like early termination and significant head motion contribute to failed scans, causing the absence of valid data in the longitudinal dataset. Therefore, completing the images of the missing time points in the longitudinal dataset is crucial as it enhances the reliability of neurodevelopmental analyses. Traditional methods involve fitting the structural or morphological information from incomplete data through a simple linear mixed-effect model. This approach often overlooks highly non-linear developmental patterns and degrades with more missing data.

Thus, a complex model that can learn the nonlinear brain developmental relationships is required for the collaborative completion of longitudinal neuroimage data. Many deep learning-based methods have been proposed to establish the mapping between existing and missing images, such as CNN and GAN-based models. However, such models always drive the result to an average of several possible missing predictions (e.g., model collapse), generating population average-like images without sufficient individual diversity. Such diversity is considered essential for development studies.

\begin{figure*}[h]
\centering
\includegraphics[width=\linewidth]{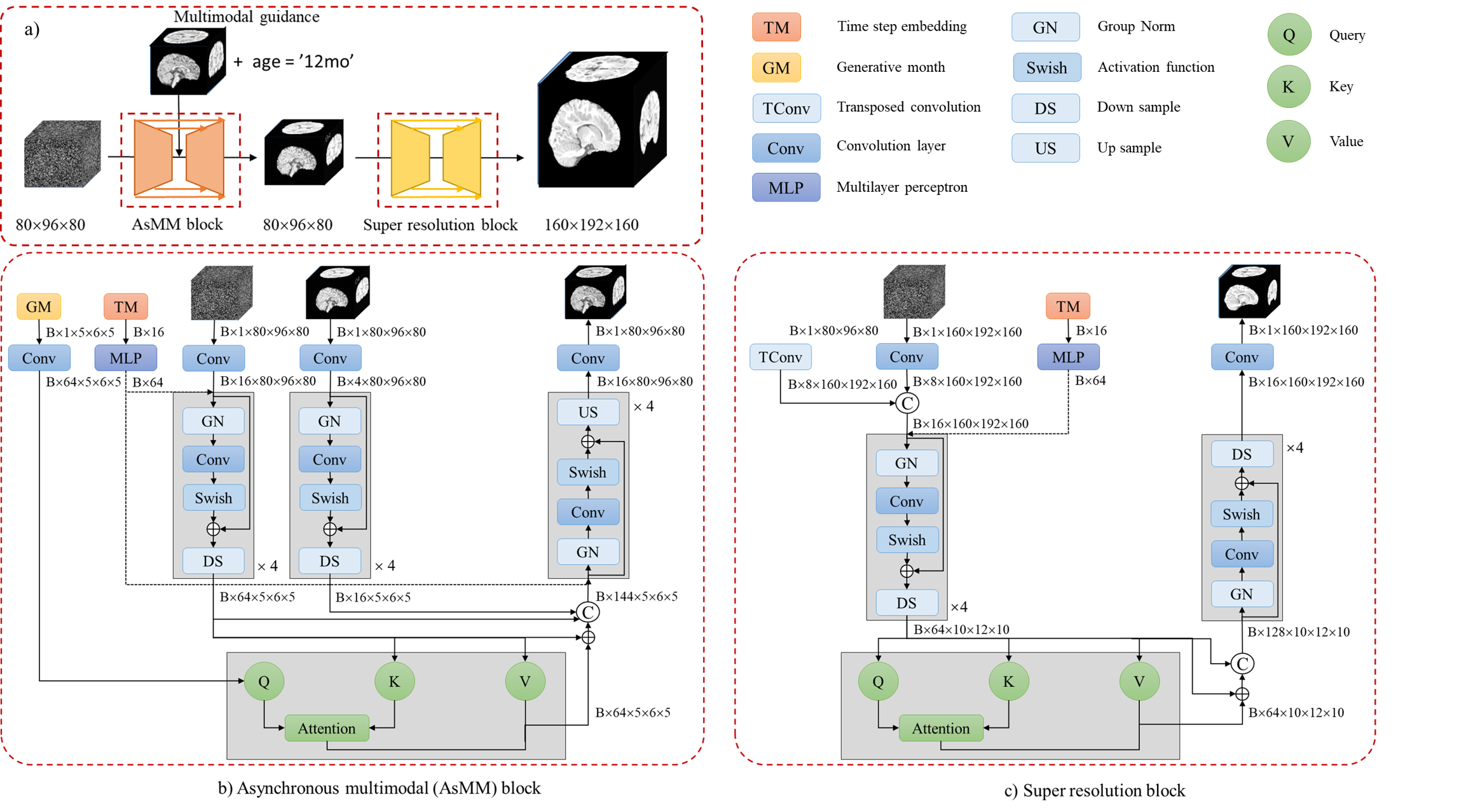} 
\caption{The denoising module in generate and fine stages. a) The cascaded diffusion pipeline for image completion consists of an image generation model and a super-resolution model. b) The asynchronous multimodal (AsMM) block. c) The super-resolution block with transposed convolution.}
\label{Fig:Model}
\end{figure*}

Recently, the denoising diffusion probabilistic model (DDPM)\cite{song2020denoising} has shown remarkable success in various tasks, such as conditioned image generation\cite{rombach2022high, meng2022novel} and video synthesis\cite{ruan2022mm}. Inspired by the DDPM, this paper proposes a novel cascaded diffusion model to effectively assure the generated images have the desired semantics. Specifically, our model firstly generates a structured image with multimodal guidance and then refines the synthesized image by a super-resolution diffusion model. An asynchronous 3D multimodal denoising block (AsMM Block) is designed to disentangle multimodal guidance and more effective use of multimodal information to generate the brain in the reverse denoising process.

\section{Methods}

DDPM is still challenging to infer a high-quality 3D brain image with huge details and neuroscientific fidelity efficiently. Hence, we develop a cascaded two-stage diffusion-based pipeline\cite{ho2022cascaded} for generating high-quality infant brain MRI at missing time points, shown in Fig \ref{Fig:Model}a. Suppose $x_0$ is the high-resolution data and $z_0$ is its low-resolution data. Under low resolution, we have a diffusion model to generate structured longitudinal images $P_c(z_0|X_{mo})$; under high resolution, we have a fine super-resolution diffusion model to refine the low-resolution image to the high-resolution image $P_f(x_0|z_0)$. The cascaded pipeline forms a latent variable model of the high-resolution data, that is, $P_c(z_0) = \int P_f(x_0|z_0) dz_0$.

\subsection{Generate Stage: Diffusion-based 3D Medical Image Completion}
The first stage shown in Fig \ref{Fig:Model}b is to generate a structured image with subject-specific guidance using diffusion model. The mathematical description of the conditional diffusion model is shown in formula \eqref{formula: conditional}.
\begin{equation}
    \begin{aligned}
        p_{\theta}(x_{0:T}|c) &= p(x_{T}) \prod^{T}_{t=1} p_{\theta}(x_{t-1}|x_{t},c)\\
        p_{\theta}(x_{t-1}|x_{t},c) &= \mathcal{N}(x_{t-1};\mu_{\theta}(x_{t},t,c), \Sigma_{\theta}(x_{t}, t,c))\label{formula: conditional}
    \end{aligned}
\end{equation}

We consider a set of paired images and age text as the multimodal guidance $c = (X_{im}, X_{mo})$ for both desired image and age information. The model training phase consists of two processes: the diffusion process and the reverse process. In the diffusion process, the low-resolution data $z_0$ sampling $t$ time steps becomes a hidden variable in the Gaussian distribution $x_t$. In reverse denoising process, $x_t$ and $X_{im}$ are encoded by independent encoders including 4 residual blocks which consist of 3D convolution layers and average pooling layer. $X_{mo}$ and $t$ are embedded by 3D convolution and MLP respectively. The structure of the decoder is similar to the encoder, except that the up-sampling uses transposed convolution. Specifically, age information $X_{mo}$ is applied to attention mechanisms, and denoising time step $t$ is applied to the residual block. In practice, since we take sample quality as the main purpose, we train with a simplified loss function in formula \eqref{loss_func}.  

\begin{equation}
    \begin{aligned}
        L^{simple} 
        &\triangleq E[||\epsilon - \epsilon_{\theta}(x_{t};t,\tau(X_{mo}), X_{im})||^{2}]
    \end{aligned}
    \label{loss_func}
\end{equation}

In addressing the entanglement issue in the multimodal guidance during the reverse denoising process, we have developed an asynchronous 3D multimodal denoising block (AsMM block) architecture. Specifically, we have employed two distinct methods to guide the denoising process with both images and age. The subject-specific image is conditioned using a skip-connection by concatenating with the denoising images. An independent encoder is used for the denoising image and conditioning image in the reverse process. The subject's age is added to the information bottleneck using attention mechanism. Then age information is summed up with the image feature via the embedding layer.

\subsection{Refine stage: Conditional 3D super-resolution}
After getting the low-resolution generated image $P_c(z_0|X_{mo})$, we further refine high-resolution generated images using a super-resolution model to add more detailed information. Fig \ref{Fig:Model}c illustrates the refined super-resolution diffusion model consisting of diffusion process and reverse process. The diffusion process gradually adds Gaussian noise to the original high-resolution image $x_0$ and becomes a hidden variable $x_t$ in the Gaussian distribution. In the reverse process, $P_c(z_0|X_{mo})$ is up-sampled by transposed convolution and used as conditions to concatenate with $x_t$ as model inputs and reconstructed to be high-resolution image $P_f(x_0|z_0)$ via denoising block. To efficiently condition the model on the low-resolution guided image, we adopt transposed convolution blocks instead of bilinear interpolation. After training, at inference stage, the low-resolution image $P_c(z_0|X_{mo})$ is up-sampled and concatenated with randomly sampled Gaussian noise to input into model to transform itself into a high-resolution image. We adopt DDIM\cite{song2020denoising} with a fine-tuning 3D U-net architecture that is trained with an objective function $E[||\epsilon - \epsilon_{\theta}(x_{t};t, X_{im})||^{2}]$ to iterative denoising at various pixel levels.

\section{Experiments and results}
\subsection{Dataset and Image Processing}
We used large-scale longitudinal brain imaging data from the Baby Connectome Project (BCP)\cite{howell2019unc}. 654 T1w MRI scans from 340 subjects aged 2 weeks to 26 months were kept for the following experiments. We selected 30 MRI scans as a test set and ensured that the test cases covered the entire age range (i.e., 2 weeks to 26 months). All T1w MRI scans were acquired using the 3T Siemens Prisma MRI scanner with the following parameters: TR/TE = 2400/2.24ms, resolution = 0.8$\times$0.8$\times$0.8 $mm^3$, with an image size of 256$\times$320$\times$256.

All structural MRIs were processed using a standard procedure followed by\cite{li2013mapping}, including skull stripping, rigid alignment to the brain atlas, and segmentation. For data augmentation, the multimodal guidance images were rotated randomly along the x, y, and z axis with a randomly selected rotation degree between 0 and 5 degrees. This augmentation was only applied to guidance images to improve the robustness of the generation model. Considering the computational cost, the input data of the first stage generation model were down-sampled to the resolution of 2$\times$2$\times$2 $mm^3$ and cropped into the size of 80$\times$96$\times$80. During the second refine stage, the training data model was re-sampled to the resolution of 1$\times$1$\times$1 $mm^3$ and resized to 160$\times$192$\times$160.

\subsection{Implementation Details}
Our model was implemented using PyTorch. Network was constructed based on DDIM\cite{song2020denoising} with the noise level set from 1e-4 to 5e-3 by linearly scheduling with 4000 steps in the generation model, and 1e-4 to 2e-2 with 1000 steps for the super-resolution stage. Adam optimization was used with a learning rate of 2e-4. The generation model was trained on two Nvidia A100 GPUs with 80 GB memory each for 5 days. The super-resolution model was trained on four Nvidia A100 GPUs for 5 days.

\subsection{Results}
\begin{figure*}[h]
\centering
\includegraphics[width=0.9\textwidth]{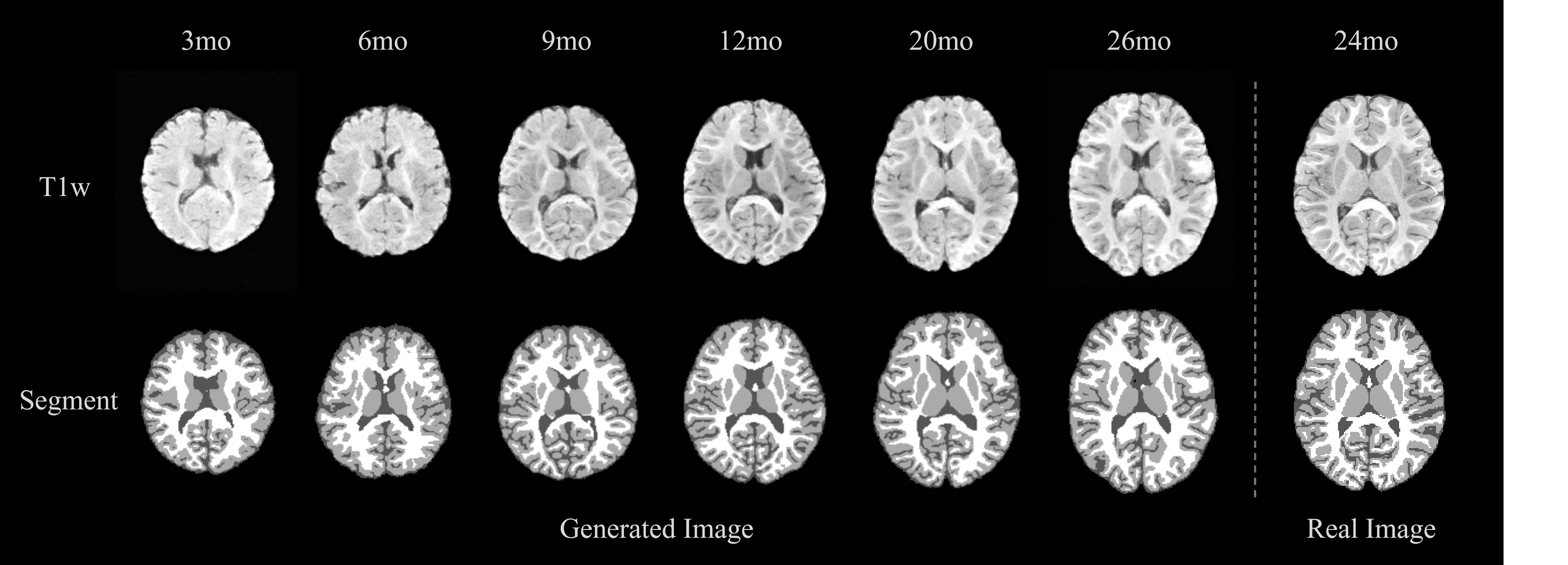} 
\caption{Showcase of the generated longitudinal infant brain images by our proposed Cas-DiffCom from the same subject. The top row displays the generated T1w MRI scans. The bottom row shows the corresponding brain segmentation results. The age time points are 3, 6, 9, 12, 20, and 26 months old. The image on the right at 24 months old is the real image from the same subject. }
\label{Fig:result}
\end{figure*}

After Cas-DiffCom was trained, the model could use guidance images and age information to generate all the MRIs for the subject at missing time points. The final results of the generated images are shown in Fig \ref{Fig:result}. Segmentation is completed by the image analysis tool, uAI Research Portal \cite{shi2022deep}\cite{zhang2018longitudinally}\cite{shi2010neonatal}. It can be seen that the details were well preserved in the generated images by our method. The volume of the brain is increasing, and the contrast between white and grey matter in the brain has significantly improved from 3 to 9 months.

\begin{figure*}[h]
\centering
\includegraphics[width=0.9\textwidth]{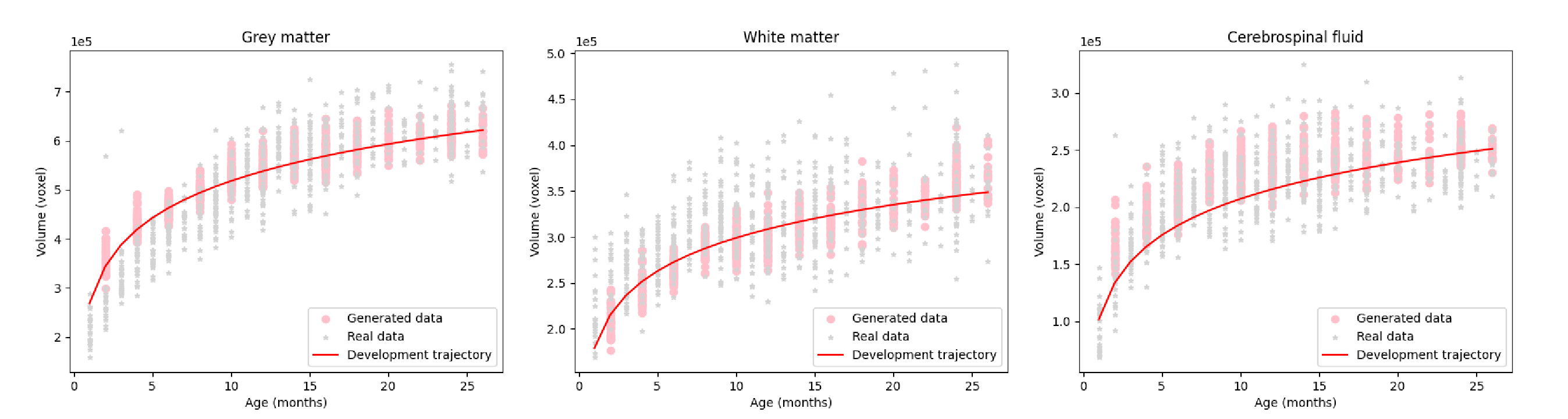} 
\caption{Comparison of brain tissue growth characterization between the generated data (red) and the ground truth (grey). The ground truth total volumes of the grey matter, white matter and cerebrospinal fluid fitted very well with the development trajectories estimated from the generated images.}
\label{Fig:curve}
\end{figure*}

We evaluate the visual quality and similarity between the real and generated images qualitatively with two widely used metrics: peak signal-to-noise ratio (PSNR) and structural similarity index measure (SSIM).
First, we chose conditional GAN as the baseline. The results illustrated that the GAN model faced difficulty on infant longitudinal brain image completion. In addition, we compared the results generated by different conditional diffusion-based networks to validate the effectiveness of the AsMM module. Model 1 is a conditional DDIM without an independent guidance image encoder. Model 2 is a conditional DDIM with the AsMM block. Table \ref{Table:1} shows the results of the comparison between the four models. The PSNR and SSIM of our proposed Cas-DiffCom are 24.15 and 0.81, respectively, which illustrate that our method performs better than all other compared models.

\begin{table}[h] 
\centering
\setlength{\tabcolsep}{2mm}
\begin{tabular}{ccccccccc} 
\hline 

\multicolumn{3}{c}{}&cGAN&Model 1&Model 2&Our Method&\\  
\hline 
\multicolumn{3}{c}{PSNR}&17.76&22.33&23.22&\textbf{24.15}&\\   
\multicolumn{3}{c}{SSIM}&0.72&0.73&0.78&\textbf{0.81}&\\
\hline 
\end{tabular}
\caption{Our method using an asynchronous block with an independent guided image encoder has the best SSIM of 0.81 and PSNR of 24.15.}
\label{Table:1}
\end{table}

To better evaluate our proposed model, we also validated its ability to complete missing data from the neuroscience perspective. Specifically, the development trajectories characterizing longitudinal growth of infant brain tissue volume were fitted with the completion data using the linear mixed-effect model with the log-linear function. In Figure \ref{Fig:curve}, the volume of brain white matter, grey matter and cerebrospinal fluid from the real data was shown with a grey scatter plot, while the generated data was shown with a pink scatter, with the fitted trajectories calculated from the complete data, including real data and generated data. The results show that the generated data closely approximates the real distribution, and the calculated trajectory from the completed data exhibits a similar trend to the real data. 

\section{Conclusion}
In this work, we present a novel cascaded diffusion model (Cas-DiffCom) with multimodal guidance for longitudinal infant brain image completion. We propose an asynchronous multimodal block for multimodal guidance disentanglement. Our refinement approach for 3D brain image super-resolution further improves the image realism and sharpness of the generated images. Our approach provides a promising solution to employ the generated missing data for downstream neuroscience tasks with high fidelity and implications.

\bibliographystyle{IEEEbib}
\bibliography{refs}

\end{document}